\documentclass[osajnl,twocolumn,showpacs,superscriptaddress,10pt]{revtex4-1}
\usepackage{amsmath,amssymb,graphicx}

\newcommand{\dk}{\text{d}k}

\newcommand{\NDFG}{\mathcal{N}^\text{DFG}_\text{sing}}
\newcommand{\NSFG}{\mathcal{N}^\text{SFG}_\text{sing}}
\newcommand{\NSPDC}{\mathcal{N}^\text{SPDC}_\text{pair}}
\newcommand{\delks}{\delta k_\text{s}}
\newcommand{\delki}{\delta k_\text{i}}
\newcommand{\delkp}{\delta k_\text{p}}

\newcommand{\ka}{k_\text{a}}
\newcommand{\kb}{k_\text{b}}
\newcommand{\ks}{k_\text{s}}
\newcommand{\ki}{k_\text{i}}
\newcommand{\kp}{k_\text{p}}
\newcommand{\Fks}{{\text{F}\ks}}
\newcommand{\Fki}{{\text{F}\ki}}
\newcommand{\SHkp}{{\text{SH}\kp}}

\newcommand{\opa}[1]{a_{#1}}
\newcommand{\opad}[1]{a_{#1}^\dagger}
\newcommand{\opnuma}[1]{\opad{#1}\opa{#1}}
\newcommand{\betaSHk}{\beta_{\text{SH}k}}
\newcommand{\betaSHkp}{\beta_{\text{SH}\kp}}
\newcommand{\betaFp}{\beta_{\text{F}+}}
\newcommand{\betaFm}{\beta_{\text{F}-}}

\newcommand{\expect}[2]{\left \langle\psi\right\vert_\text{#1} #2 \left\vert\psi \right \rangle_\text{#1}}

\begin{document}

\title{The effect of scattering loss on connections between classical and quantum processes in second-order nonlinear waveguides}

\author{L. G. Helt}

\affiliation{Centre for Ultrahigh bandwidth Devices for Optical Systems (CUDOS), QSciTech Research Centre, MQ Photonics Research Centre, Department of Physics and Astronomy,
Macquarie University, NSW 2109, Australia}

\author{M. J. Steel}

\affiliation{Centre for Ultrahigh bandwidth Devices for Optical Systems (CUDOS), QSciTech Research Centre, MQ Photonics Research Centre, Department of Physics and Astronomy,
Macquarie University, NSW 2109, Australia}

\begin{abstract}We show that a useful connection exists between spontaneous parametric downconversion (SPDC) and sum frequency generation in nonlinear optical waveguides with arbitrary scattering loss, while the same does not hold true for SPDC and difference frequency generation. This result deepens the relationship between quantum and classical second-order nonlinear optical processes in waveguides, and identifies the most accurate characterization of their quantum performance in the presence of loss based solely on classical measurements.\end{abstract}


\maketitle

While quantum field theory is required to fully explain the interaction of light with matter, the standard description of nonlinear optical interactions is grounded in classical electromagnetic theory. For example, sum frequency generation (SFG) can be thought of as a kind of photon fusion that proceeds efficiently if the momenta of the three photons involved are appropriately
matched.  However, it is more commonly viewed as a three-wave mixing process that requires phase matching conditions be met in order to be efficient. In fact, the two pictures are deeply connected~\cite{Kleinman:1968}, and allow the interpretation of quantum nonlinear optical processes as the spontaneous counterparts of corresponding classical processes.

This relationship has recently become of practical importance for characterizing photon sources based on spontaneous quantum nonlinear optical processes. The key realization is that the biphoton wave function for a quantum source is precisely the response function for a classical field generated by the same device and input field, with an appropriate additional input seed field~\cite{Liscidini:2013}. This allows direct but slow coincidence detection measurements of a spontaneous process to be supplanted by faster and more convenient optical power measurements of the corresponding classical process. Already, the new measurement technique has enabled previously unobtainable resolution in the spectral characterization of two-photon states from various waveguides including an \mbox{AlGaAs} ridge~\cite{Eckstein:2014}, optical fibers~\cite{Fang:2014}, and a silicon nanowire~\cite{Jizan:2014}.  It has also facilitated a simple and efficient polarization density matrix reconstruction method~\cite{Liscidini:2014}.  The technique will only become more valuable as improvements in fabrication technology lead to more and more devices on the same chip. 

However, theoretical results and experimental analysis thus far have relied on calculations performed in the limit of zero photon loss. As integration and miniaturization of photon sources continues, this assumption will become fragile, since attenuation mechanisms such as Rayleigh scattering by waveguide impurities, or scattering due to inevitable sidewall roughness~\cite{Melati:2014}, will become more significant.  With waveguide SPDC sources becoming more widely used~\cite{Fiorentino:2007,Fujii:2007,Horn:2012,Zhong:2012,Orieux:2013, Harder:2013,Kaiser:2014,Solntsev:2014}, we should understand how to best characterize them under realistic scenarios.

\begin{figure}[hbt]
\includegraphics[width=0.5\textwidth]{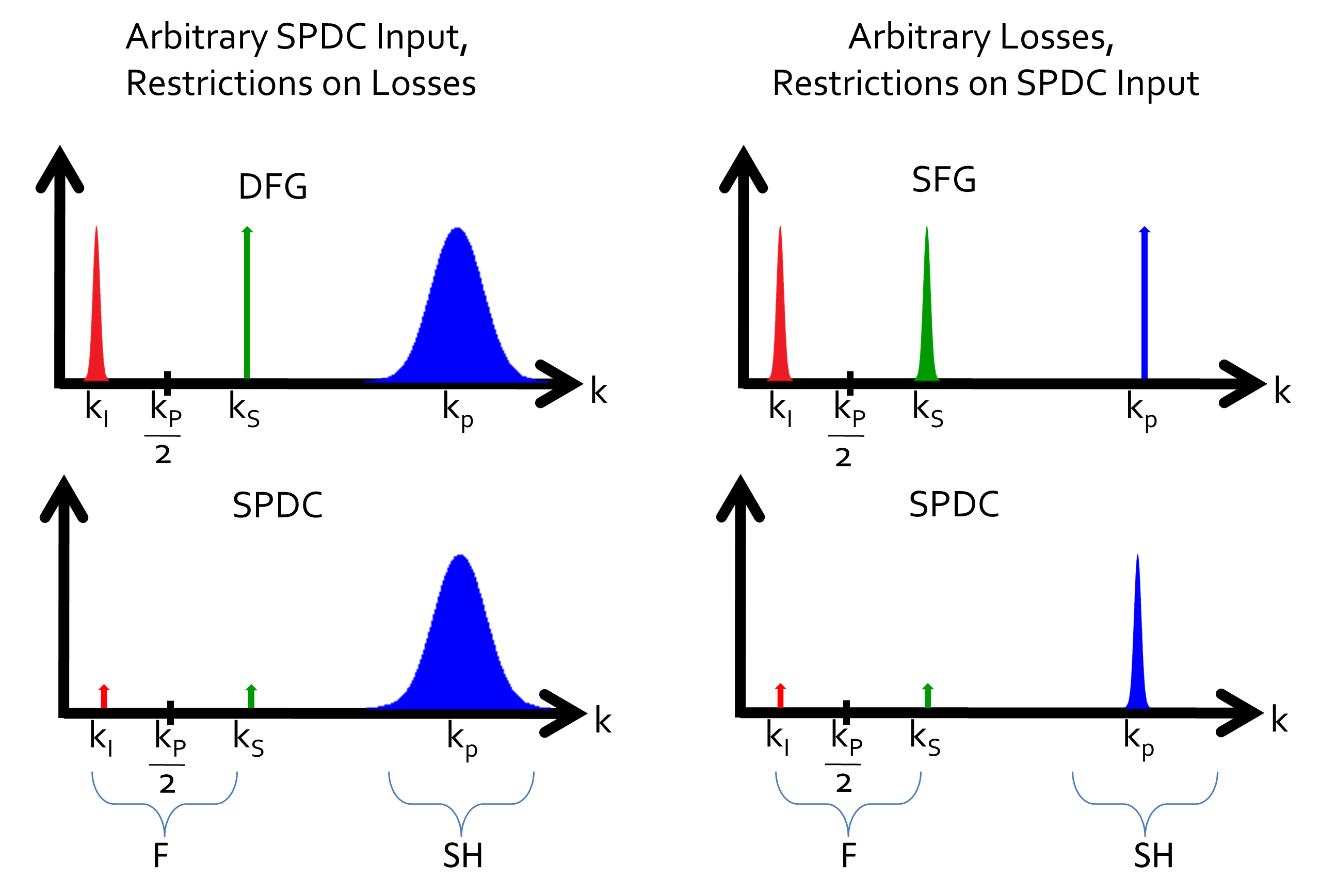}
\caption{\label{fig:sketch} Sketch of connections between input fields and generated fields, Gaussians and arrows, respectively, for quantum and classical second-order nonlinear optical processes.}
\end{figure}

In this letter we discuss the link between spontaneous parametric downconversion (SPDC) and two classical three-wave mixing processes in waveguides, and how this link is affected by scattering loss. We focus on second-order nonlinear ($\chi^{(2)}$) processes for simplicity, but expect similar arguments to carry over to third-order processes as well---a topic we reserve for future work. As all $\chi^{(2)}$ processes in a given device are governed by the same susceptibility, knowledge of any particular process yields at least some information about others. Dividing the spectrum into ``fundamental'' (F) and ``second harmonic'' (SH) frequency bands, one might argue that difference frequency generation (DFG) is the most faithful analog of SPDC because, as illustrated in Fig.~\ref{fig:sketch}, the two processes may share identical input fields in the SH band. In contrast, for SFG the generated field, rather than an input field, is the proxy for the SPDC input. However, in the presence of loss, this advantage of DFG over SFG does not persist in general. In particular, we show that a DFG experiment only gives information about an SPDC experiment in the same device for arbitrary SPDC input fields if losses in the F band are either small or approximately equal to losses in the SH band. Alternatively,  an SFG experiment gives useful information for arbitrary losses provided the SPDC input fields of interest are quasi-continuous wave (CW) (see Fig.~\ref{fig:sketch}).  The difference between DFG and SFG arises because the temporal evolution of input fields in the presence of loss is different to that of generated fields in the presence of loss~\cite{Sutherland:2003}.

Our strategy to obtain these results is  to describe both classical and quantum nonlinear optical processes within a common Hamiltonian framework.  This requires the more involved apparatus of quantum states and operators for the calculation of classical processes that would normally be treated with standard coupled mode equations \cite{Sutherland:2003}. However, it keeps notation consistent and allows a direct comparison between classical and quantum processes.  In this picture, classical processes appear as transformations of input coherent states to generated coherent states of different wavelengths; the average number of photons in each coherent state is directly proportional to the more commonly calculated power~\cite{Helt:2013thesis}.

As in \cite{Helt:2014}, we simultaneously treat scattering loss and a nonlinear optical process by extending the backward Heisenberg picture approach of Yang et al.\ \cite{Yang:2008} to include the interaction of guided waveguide modes with reservoir modes. Our full Hamiltonian is $H=H_{\text{L}}+H_{\text{NL}}+H_{\text{R}}+H_{\text{C}}$, where the linear optical, nonlinear optical, reservoir, and coupling pieces are respectively
\begin{align}
&H_{\text{L}}=\sum_{m=\text{F,SH}}\int \dk\,\hbar \omega_{mk}a_{mk}^{\dagger }a_{mk},\nonumber\\
&H_{\text{NL}}=-\int \dk_{1}\dk_{2}\dk\,S\left(k_{1},k_{2},k\right) a_{\text{F}k_{1}}^{\dagger }a_{\text{F}k_{2}}^{\dagger}a_{\text{SH}k}+\text{H.c.},\nonumber\\
&H_{\text{R}}=\sum_{m=\text{F,SH}}\int \dk\text{d}\mu\,\hbar \Omega_{m\mu k}b_{m\mu k}^{\dagger }b_{m\mu k},\nonumber\\
&H_{\text{C}}=\sum_{m=\text{F,SH}}\int \dk\text{d}\mu\,\hbar c_{m\mu k}a_{mk}^{\dagger }b_{m\mu k}+\text{H.c.},
\end{align}
and H.c.\ denotes Hermitian conjugate. Following Fig.~\ref{fig:sketch}, we label the modes by $m=\text{F}$ for fundamental and $m=\text{SH}$ for second-harmonic. Additionally, $\mu $ is a shorthand for all quantities necessary to specify reservoir modes in addition to $m$ and $k$, and $c_{m\mu k}$ are waveguide-reservoir coupling coefficients. The nonlinear parameters (effective area, nonlinear susceptibility, etc.)  reside in the real and positive coupling term $S$. Its definition can be found in either \cite{Yang:2008} or \cite{Helt:2014} but is unnecessary for our present discussion. 

A subtlety of the classical calculations compared to the SPDC calculation is that the output state of the waveguide contains non-commuting operators and does not immediately factor into a pump field state and a generated field state. Focusing only on the process at hand (e.g. neglecting SPDC and second-harmonic generation contributions in the DFG calculation) we use the Baker-Campbell-Hausdorf relation \cite{Kim:1991} to keep only leading-order terms in $S$ and separate the calculated output state into a tensor product of a state containing the input field(s) and a state containing the generated field. Following this separation, we form density operators for the states of generated photons and trace over the reservoir operators at zero temperature. For classical processes the resulting reduced density operator is pure, as a damped coherent state remains coherent throughout its evolution~\cite{Carmichael:1999}, and so we simply give kets below. 

We present all of our results in $k$-space for notational convenience and for ease of comparison with earlier work~\cite{Liscidini:2013}. However note that for many devices it is quite reasonable to think of $k$ as a proxy for $\omega$, approximating $\dk\rightarrow\text{d}\omega\,\dk/\text{d}\omega\approx\text{d}\omega\,v_\text{g}^{-1}$ where $v_\text{g}$ is a group velocity constant over the frequency range of interest~\cite{Yang:2008}. There are a number of fields to be distinguished which we label by their center wavenumbers. In deriving our general equations we use $\ka$, $\kb$ as generic labels for fields in either band. In identifying the quantum-classical correspondence for particular scenarios, we use $\ks$ (``signal'') and $\ki$ (``idler'') for the F band, and $\kp$ (``pump'') for the SH band (see Fig.~\ref{fig:sketch}).

We first consider the DFG input state
\begin{equation}
\left\vert \psi \right\rangle _{\text{in}}=e^{ \left( \int \dk\left(z_{\text{F}k_{\text{a}}}\phi _{\text{F}k_{\text{a}}}\left( k\right) a_{\text{F}k}^{\dagger }+z_{\text{SH}k_{\text{b}}}\phi _{\text{SH}k_{\text{b}}}\left(k\right) a_{\text{SH}k}^{\dagger }\right) -\text{H.c.}\right)} \left\vert \text{vac}\right\rangle ,
\end{equation}
which describes the tensor product of two general coherent states containing an average of $\left\vert z_{\text{F}k_{\text{a}}}\right\vert ^{2}$ and $\left\vert z_{\text{SH}k_{\text{b}}}\right\vert ^{2}$ photons, with input pulse waveforms centered at $k_{\text{a}}$ and $k_{\text{b}}$, respectively, that satisfy $\int $d$k\left\vert\phi _{mk_n}\left( k\right) \right\vert ^{2}=1$. For this input state we find the generated state
\begin{equation}\label{eq:cohdfg}
\left\vert \psi \right\rangle _{\text{DFG}}=e^{ \left( \int_{t_{0}}^{t_{1}}\text{d}t\int \dk\,\Phi _{\text{DFG}}\left( k;t\right) e^{-\beta _{\text{F}k}\left( t_{1}-t\right) }a_{\text{F}k}^{\dagger }-\text{H.c.}\right)}\left\vert \text{vac}\right\rangle ,
\end{equation}
where
\begin{align}
\Phi _{\text{DFG}}\left( k;t\right) =&\frac{2iz_{\text{F}k_{\text{a}}}^{\ast}z_{\text{SH}k_{\text{b}}}}{\hbar }\int \dk_{1}\dk_{2}\,\phi _{\text{F}k_{\text{a}}}^{\ast }\left( k_{1}\right) \phi _{\text{SH}k_{\text{b}}}\left( k_{2}\right)\nonumber\\
&\times S\left( k,k_{1},k_{2};t\right) e^{-\left( \beta _{\text{F}k_{1}}+\beta _{\text{SH}k_{2}}\right) \left( t-t_{0}\right) },
\end{align}
and we have defined
\begin{equation}
S\left( k_{1},k_{2},k;t\right) =S\left( k_{1},k_{2},k\right) e^{i\left(\omega _{\text{F}k_{1}}+\omega _{\text{F}k_{2}}-\omega _{\text{SH}k}\right)t}  \label{S}.
\end{equation}
The loss rates $\beta _{mk}$ are related to the coefficients in the coupling Hamiltonian via $\beta _{mk}=\mathcal{C}_{mk}\pi \mathcal{D}_{mk}>0$, with $\left\vert c_{m\mu k}\right\vert ^{2}\approx \mathcal{C}_{mk}$, d$\mu $/d$\Omega _{m\mu k}\approx \mathcal{D}_{mk}$, as well as to the usual attenuation coefficients $\alpha _{mk}$ via $\alpha _{mk}=2\beta _{mk}/v_{m}$ where $v_{m}$ are group velocities \cite{Helt:2014}. The times $t_0$ and $t_1$ refer to the instants at which the nonlinear interaction starts and stops, respectively, and as a first approximation can be taken such that $t_1-t_0=L/v_\text{in}$, where $v_\text{in}$ is the largest input field group velocity and $L$ the length of the waveguide. In an obvious notation, an SFG calculation carried out in a similar manner yields
\begin{equation}\label{eq:cohsfg}
\left\vert \psi \right\rangle _{\text{SFG}}=e^{ \left( \int_{t_{0}}^{t_{1}}\text{d}t\int \dk\,\Phi _{\text{SFG}}\left( k;t\right) e^{-\beta _{\text{SH}k}\left( t_{1}-t\right) }a_{\text{SH}k}^{\dagger }-\text{H.c.}\right)}\left\vert \text{vac}\right\rangle,
\end{equation}
where
\begin{align}
\Phi _{\text{SFG}}\left( k;t\right) =&\frac{2iz_{\text{F}k_{\text{a}}}z_{\text{F}k_{\text{b}}}}{\hbar }\int \dk_{1}\dk_{2}\,\phi _{\text{F}k_{\text{a}}}\left( k_{1}\right) \phi _{\text{F}k_{\text{b}}}\left(k_{2}\right)\nonumber\\
&\times S^{\ast }\left( k_{1},k_{2},k;t\right) e^{-\left( \beta _{\text{F}k_{1}}+\beta _{\text{F}k_{2}}\right) \left( t-t_{0}\right) }. \label{SFGdef}
\end{align}
Note that here, both $\ka$ and $\kb$ index fields in the F band. We stress that Eqs.~\eqref{eq:cohdfg} and~\eqref{eq:cohsfg} are familiar results recast in an unusual form. For example, the SH field in \eqref{eq:cohsfg} is generated by the two F input field waveforms in \eqref{SFGdef}.

We now turn to the quantum process. Previously we have shown that, in the limit of a low probability of pair production, the reduced density operator describing generated SPDC photons consists of two-photon, one-photon, and vacuum pieces, with the one photon contribution being a mixed state~\cite{Helt:2014}. However, only the two-photon piece can lead to non-zero coincidence detection, and it can be written as a pure state with the ket
\begin{align}
\left\vert \psi \right\rangle _{\text{II}}=&\frac{1}{\sqrt{2}}\int_{t_{0}}^{t_{1}}\text{d}t \int \dk_{1}\dk_{2}\,\phi \left(k_{1},k_{2};t\right)\nonumber\\
&\times e^{-\left( \beta _{\text{F}k_{1}}+\beta _{\text{F}k_{2}}\right) \left( t_{1}-t\right) }a_{\text{F}k_{1}}^{\dagger }a_{\text{F}k_{2}}^{\dagger }\left\vert \text{vac}\right\rangle ,
\end{align}
where
\begin{align}
\phi \left( k_{1},k_{2};t\right) =&\frac{\sqrt{2}iz_{\text{SH}k_{\text{a}}}}{\hbar }\int \dk\phi _{\text{SH}\ka}\left( k\right)\nonumber\\
&\times S\left(
k_{1},k_{2},k;t\right) e^{-\beta _{\text{SH}k}\left( t-t_{0}\right) }.
\end{align}

Suppose we denote as $\NSPDC \delks \delki =\expect{SPDC}{\opnuma{\Fks} \opnuma{\Fki} } \delks\delki$, the average number of SPDC photon pairs per pump pulse produced over
bandwidths $\delks$, $\delki$. Likewise, $\NDFG \delki = \expect{DFG}{\opnuma{\Fki} } \delki$, is  the average number of photons per pulse produced over $\delki$ by a DFG process with
the same SH input field, $\phi_{\text{SH}\kp}\left(k\right)$, as well as a second spectrally narrow input field $\phi_{\text{F}\ka}\left(k\right)$ centered at $\ka=\ks$~\cite{Liscidini:2013}. By ``spectrally narrow'' we mean  narrower than any features in $S$ or the loss rates. Then in the absence of loss, one can show that the number ratio of classical photons and quantum pairs satisfies~\cite{Liscidini:2013}
\begin{align}
\label{eq:rdfg1}
R^\text{DFG}\equiv\frac{\NDFG \delki }{\NSPDC \delks \delki} = |z_\Fks|^2.
\end{align}
However, with scattering loss included, and taking $t_{1}=-t_{0}=T$ as well as $v_\text{F}/v_\text{SH}\approx 1$, the ratio becomes
\begin{align}
\label{eq:rdfg2}
R^\text{DFG}
\approx |z_\Fks|^2 \frac{I[\betaFm-\betaSHk]}{I[\betaFp-\betaSHk]},
\end{align}
where
\begin{align}\label{eq:idef}
I[x]=
\left\vert \int_{-T}^{T}\text{d}t\int \dk\, \phi _{\text{SH}\kp}(k) S\left(\ks,\ki,k;t\right) e^{xt }\right\vert ^{2}
\end{align}
and $\beta_{\text{F}\pm}=\beta_\Fki\pm\beta_\Fks$.  

Similarly, we can consider the ratio between the average number of SPDC pairs generated over $\delks$, $\delki$, and single photons generated over $\delkp$ for an SFG process with spectrally narrow input fields $\phi_{\text{F}\ka}\left(k\right)$ and $\phi_{\text{F}\kb}\left(k\right)$ centered at $\ka=\ks$ and $\kb=\ki$, respectively.  Defining
$\NSFG \delkp  =\expect{SFG}{\opnuma{\SHkp}  } \delkp$, we find
\begin{align}
\label{eq:rsfg1}
R^\text{SFG} & \equiv\frac{\NSFG \delkp }{\NSPDC \delks \delki} \nonumber \\
&\approx \frac{|z_\Fks|^2 |z_\Fki|^2}{ |z_{\text{SH}\kp}|^2} \frac{\left\vert \int_{-T}^{T}\text{d}t\,  S\left(\ks,\ki,\kp;t\right) e^{\left(\betaFp -\betaSHk\right)t}\right\vert ^{2}
}{   I[\betaFp-\betaSHk]},
\end{align}
where we  have used the property $S^{\ast }\left(k_{1},k_{2},k;-t\right) =S\left( k_{1},k_{2},k;t\right) $ (recall \eqref{S}). 

In general, one does not know the details of the physical parameters inside the nonlinear coupling $S$, so for Eqs.~\eqref{eq:rdfg2} or~\eqref{eq:rsfg1} to be useful, the ratio of integrals
needs to reduce to unity. While a spectrally narrow $\phi _{\text{SH}\kp}\left( k\right)$ will not help to simplify $R^\text{DFG}$, this is not the case for $R^\text{SFG}$. For SFG photon production over $\delta k_\text{p}$ corresponding to SPDC with a spectrally narrow input field $\phi _{\text{SH}\kp}\left( k\right)$ centered at $\kp$, Eq.~\eqref{eq:idef} shows that the second fraction in Eq.~\eqref{eq:rsfg1} tends to unity and 
\begin{equation}
\label{eq:rsfg2}
R^\text{SFG} \approx\frac{\left\vert z_{\text{F}k_{\text{s}}}\right\vert ^{2}\left\vert z_{\text{F}k_{\text{i}}}\right\vert ^{2}}{\left\vert z_{\text{SH}k_{\text{p}}}\right\vert ^{2}}.
\end{equation}
These results illustrate that in order to exploit the quantum-classical correspondence for nonlinear optical processes, one of two conditions must hold: If scattering losses in the F band are small or equal to those in the SH band, then Eq.~\eqref{eq:rdfg2} reduces to~\eqref{eq:rdfg1} and the expected performance of the device in an SPDC experiment can be obtained from a DFG experiment. Alternatively, if scattering losses in the F and SH bands are unconstrained but the input field used for SPDC is narrow in $k$ space (i.e. quasi-CW)~\cite{Yang:2008}, then the connection between SPDC and DFG is lost but Eq.~\eqref{eq:rsfg2} holds, and the expected performance can be obtained from an SFG experiment. 

\begin{figure}[hbt]
\hspace{-5mm}\includegraphics[width=0.45\textwidth]{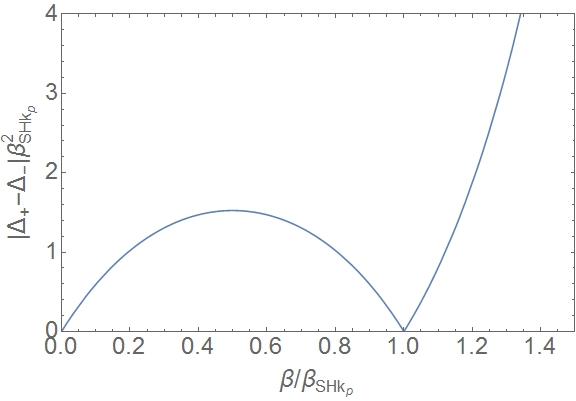}
\caption{Absolute value of the difference between the most relevant contributions to the average number of photons generated in a DFG process and an SFG process in the same waveguide for spectrally narrow input and generated fields centered at the $k$ values for which energy and momentum conservation are met exactly.} \label{fig:beta}
\end{figure}

Fundamentally, the DFG result  differs from the SFG result because $\NDFG $ involves the difference of fundamental field loss terms while $\NSFG $ involves their sum. To see this in more detail, we consider spectrally narrow input and generated fields centered at the $k$ values for which energy and momentum conservation are met exactly. This allows the maximum possible difference between the ratios of integrals appearing in $R^\text{DFG}$ and $R^\text{SFG}$, as far from these conditions there are no photons to collect anyway, regardless of the impact of scattering loss. In this limit the temporal integrals over exponentials of loss rates appearing in $R^\text{DFG}$ and $R^\text{SFG}$ are
\begin{equation}
\Delta_{\pm}\equiv\left|\int_{-T}^{T}\text{d}t\, e^{\left(\beta_{\text{F}\pm}-\beta_{\text{SH}k_{\text{p}}}\right)t}\right|^{2}=\frac{4\sinh^{2}\left[\left(\beta_{\text{F}\pm}-\beta_{\text{SH}k_{\text{p}}}\right)T\right]}{\left(\beta_{\text{F}\pm}-\beta_{\text{SH}k_{\text{p}}}\right)^{2}}.
\end{equation}
In Fig.\ \ref{fig:beta} we plot the absolute value of the difference between these two values multiplied by $\beta_{\text{SH}k_p}^{2}$ as a function of $\beta/\beta_{\text{SH}k_p}$, where $\beta_{\text{F}k_s}=\beta_{\text{F}k_i}\equiv\beta$, for $\beta_{\text{SH}k_p}T=1$. Note how they agree exactly, and thus $R^\text{DFG}$ reduces to Eq. \eqref{eq:rdfg1}, only at $\beta=0$ or $\beta=\betaSHkp$.  Their difference reaches a local maximum at $\beta=\betaSHkp/2$ and grows exponentially beyond $\betaSHkp$. The local maximum decreases with increasing $\beta_{\text{SH}k_\text{p}}$, as large SH losses dominate the behavior of both $\Delta_-$ and $\Delta_+$ for $0<\beta<\betaSHkp$.

In conclusion, we have shown that although the presence of scattering loss at fundamental frequencies spoils the connection between SPDC and DFG the same is not true for SPDC with a spectrally narrow input field and SFG.  This result enables the characterization of the quantum performance of a nonlinear optical device based solely on classical measurements whether scattering loss is present or not.

This work was supported by the ARC Centre for Ultrahigh bandwidth Devices for Optical Systems (CUDOS) (Project No. CE110001018). L. G. H. acknowledges support from a Macquarie University Research Fellowship.

\bibliography{SET_loss}

\end{document}